\documentclass[twocolumn,
nofootinbib,
prd,aps,tightenlines]{revtex4}
\usepackage{bm}
\usepackage{amssymb}

\def\nn{ \nonumber \\ }
\def\ket#1{ \left| #1 \right\rangle }

\def\bra#1{ \left\langle #1 \right| }

\def\abs#1{ \left| #1 \right| }

\def\me#1#2#3{ \left\langle #1 \left| #2 \right| #3 \right\rangle }
\def\ip#1#2{ \left\langle #1 \left| #2 \right.  \right\rangle }

\def\dq{ \mathfrak{D} }
\def\nud{\mathfrak{N}_q^{(ud)}}
\def\dqh{ D }
\def\opa{ \mathfrak{a} }
\def\opq{ \mathfrak{q} }
\def\opu{ \mathfrak{u} }
\def\opd{ \mathfrak{d} }
\def\opA{ \mathfrak{A} }
\def\sing{ \mathcal{S} }
\def\avar{b}

\begin{document}

\title{Bosonic Operator Methods for the Quark Model}

\author{Aneesh V.~Manohar}
\affiliation{Department of Physics, University of California at San Diego,
  La Jolla, CA 92093\vspace{4pt} }

\date{\today}
\begin{abstract}
Quark model matrix elements can be computed using bosonic operators and the holomorphic representation for the harmonic oscillator. The technique is illustrated for normal and exotic baryons for an arbitrary number of colors. The computations are much simpler than those using conventional quark model wavefunctions.
\end{abstract}
\maketitle

The non-relativistic quark model has been a useful tool in the study of hadrons. Baryons and mesons are described by quantum mechanical wavefunctions for non-relativistic constituent quarks. The lowest lying baryons, the $\mathbf{8}_{1/2}$ and $\mathbf{10}_{3/2}$, are three quark states with wavefunctions which are completely antisymmetric in color, and completely symmetric in position and spin-flavor.

The properties of baryons can be computed in a systematic expansion in powers of $1/N_c$, where $N_c=3$ is the number of colors in QCD. There is an exact $SU(6)$ symmetry for baryons in the $N_c \to \infty$ limit of QCD, which is broken by $1/N_c$ corrections~\cite{Largenspinflavor}. The systematic application of the $1/N_c$ expansion requires computing properties of $SU(6)$ baryon representations~\cite{DJM1,largena,largenb,DJM2,largenrefs,seventya,seventyb,seventyc,seventyd,heavybaryons}. The non-relativistic quark model also has $SU(6)$ symmetry, and is a convenient way to keep track of $SU(6)$ group theory.

Applications of the quark model to compute baryon matrix elements, and to compute $SU(6)$ Clebsch-Gordan coefficients, require evaluting quark operator matrix elements between baryon states.  It is useful to have an efficient means of computing these quark model matrix elements. The traditional approach has been to write down color $\otimes$ spin $\otimes$ flavor wavefunctions, and use them to compute matrix elements~\cite{qmbook1,qmbook2}.  The wavefunction is completely antisymmetric in color, with the quark color indices contracted using $\epsilon_{abc}$. As a result, the three quarks must enter as one of each of the three colors, and the wavefunction is written by ordering the quarks by their color index.  As an example, the proton ($J_z=1/2$) wavefunction is
\begin{eqnarray}
\ket{p} &=\frac 1 {\sqrt{18}}\left[ uud \right] \otimes \left[2 \uparrow\uparrow\downarrow - \uparrow\downarrow\uparrow-\downarrow\uparrow\uparrow\right] + \text{cyclic},
\label{1}
\end{eqnarray}
and the position of the quark in $[uud]$ is the value of its color index.
These wavefunctions are then used to compute operator matrix elements. For example, the matrix element 
\begin{eqnarray}
\me{p}{ q^\dagger \sigma^3 \tau^3 q}{p}&=&\bra{p}
 \frac 1 {\sqrt{18}}\Bigl[ \left(1+1+1 \right)\left[ uud \right] \otimes \left[2 \uparrow\uparrow\downarrow \right]\nn
&& - \left(1-1-1\right)\left[ uud \right] \otimes\left[ \uparrow\downarrow\uparrow\right]\nn
&&- \left(-1+1-1\right)\left[ uud \right] \otimes\left[ \downarrow\uparrow\uparrow\right] + \text{cyclic}\Bigr],\nn
&=& \frac{10}{18} + \text{cyclic} = \frac 5 3,
\label{4}
\end{eqnarray}
where the terms in each parenthesis are the values of $ q^\dagger \sigma^3 \tau^3 q$ for each of the three quarks. The above matrix element shows that in the quark model, the proton axial charge is $(5/3) g_A$, where $g_A = 0.75$ is the axial charge of a constituent quark~\cite{georgi}. Computations using quark model wavefunctions rapidly get complicated, particularly if one is interested in studying matrix elements for arbitrary $N_c$, or for exotic baryons, both of which involve states with more than three particles.

There is an alternate way to compute matrix elements which is more efficient, and more easily implemented for computer calculations. In the ground state baryons, and for exotic baryons, one is interested in hadron states which are completely symmetric in spin $\otimes$ flavor. In this case, one can treat the quarks as bosons, completely ignoring the color quantum numbers. One can then use bosonic operator methods developed for the harmonic oscillator~\cite{ho1,ho2} to compute the matrix elements of color singlet quark operators, since they do not care about the color of individual quarks. The bosonic method has been used previously to study the properties of large-$N_c$ baryons~\cite{DJM1,DJM2,collins}, and the relation between the quark model and chiral soliton model~\cite{Manohar:1984}.  Some of the techniques used here are related to those in Ref.~\cite{collins}, where matrix elements in the $N_c \to \infty$ limit were obtained using spin-flavor coherent states.

The one-dimensional harmonic oscillator can be quantized using creation and annihilation operators $\opa^\dagger$ and $\opa$ which satisfy the commutation relation $\left[\opa,\opa^\dagger\right]=1$. In the holomorphic representation for the harmonic oscillator~\cite{ho1,ho2}, states are given as holomorphic functions $\psi(z)$ of a complex variable $z$, $\opa^\dagger=z$, and $\opa={\rm d}/{\rm d}z$. A state
\begin{eqnarray}
\ket{\psi} &=& \sum_{n=0}^\infty c_n \ket{n} = \sum_{n=0}^\infty {c_n \over \sqrt{n!}} (\opa^\dagger)^n \ket{0}
\label{5}
\end{eqnarray}
in the occupation number basis corresponds to the function
\begin{eqnarray}
\psi(z) &=& \sum_{n=0}^\infty  \psi_n z^n = \sum_{n=0}^\infty {c_n \over \sqrt{n!}} z^n
\label{6}
\end{eqnarray}
in the holomorphic basis, with expansion coefficients $\psi_n=c_n/\sqrt{n!}$. The inner product of two states is given by
\begin{eqnarray}
\ip{\chi}{\psi}\! &=& \! \sum_{n=0}^\infty n!\, \chi_n^* \psi_n = \int {{\rm d}^2 z \over 2\pi i}
e^{- \abs{z}^2} \left[\chi(z)\right]^* \psi(z),
\label{7}
\end{eqnarray}
where ${\rm d}^2 z/(2\pi i) \equiv {\rm d}\text{Re}\, z \, {\rm d} \text{Im}\, z$. The operator ${\rm d}/{\rm d}z$ is the hermitian conjugate of $z$ using the inner product Eq.~(\ref{7}), as can be verified by integration by parts.

The quark model bosonic algebra needed to study ground state and exotic baryons~\cite{JM} is generated by quark and antiquark creation and annihilation operators $\opq^\dagger_{\imath \alpha}, \opq^{\imath \alpha}, \bar \opq^{\dagger {\imath \alpha}}, \bar \opq_{\imath \alpha}$, where $\imath=1,2$ is a spin index and $\alpha=1,\ldots,F$ is a flavor index. In the holomorphic representation, $\opq^\dagger_{\imath \alpha}$ and $ \bar \opq^{\dagger {\imath \alpha}}$ are multiplication by the complex variables $Q_{\imath \alpha}$ and $ \bar Q^{ {\imath \alpha}}$, and
$\opq^{\imath \alpha}, \bar \opq_{\imath \alpha}$ are $\partial/\partial Q_{\imath \alpha},
\partial/\partial \bar Q^{ {\imath \alpha}}$, respectively. Baryon wavefunctions are holomorphic functions $\psi\left(Q_{\imath \alpha},\bar Q^{ {\imath \alpha}}\right)$. Ground state baryons, which do not contain antiquarks, are functions $\psi\left(Q_{\imath \alpha}\right)$.  An alternative notation which is used also is $Q_{11}=u_\uparrow$, $Q_{21}=u_\downarrow$, $Q_{12}=d_\uparrow$, $\bar Q^{11}  = \bar u^\uparrow$, $\opq^{11} = \opu_\uparrow$, etc.\footnote{ $Q$ and $\bar Q$ are independent complex variables. The complex conjugate of $Q$ is $Q^*$. $\bar Q$ is in the conjugate representation to $Q$, so that $\bar u^\downarrow=\bar u_\uparrow$ is a spin-up $\bar u$, and $-\bar u^\uparrow=\bar u_\downarrow$ is a spin down $\bar u$.} Quark matrix elements can be computed using bosonic operators or the holomorphic representation, and the two methods are equivalent. The holomorphic representation is particularly convenient for computer implementation.

Quark operators are differential operators in the holomorphic representation. For example 
\begin{eqnarray}
\opq^\dagger \sigma^3 \tau^3 \opq &=& u_\uparrow {\partial \over \partial u_\uparrow}
-u_\downarrow {\partial \over \partial u_\downarrow}
-d_\uparrow {\partial \over \partial d_\uparrow}
+d_\downarrow {\partial \over \partial d_\downarrow}.
\label{10}
\end{eqnarray}
Baryon wavefunctions are simpler than in the traditional approach. The wavefunctions of the $J_z=1/2$ octet and $J_z=3/2$ decuplet baryons with $I=I_3$ are:
\renewcommand{\arraystretch}{2}
\begin{eqnarray}
\begin{array}{rclrcl}
\psi_{p} &=& \frac 1 {\sqrt 3} u_\uparrow \left[ u_\uparrow d_\downarrow - u_\downarrow d_\uparrow\right], & 
\psi_{\Delta^{++}} &=& \frac 1 {\sqrt 6} u_\uparrow^3\\
\psi_{\Sigma^+} &=& \frac 1 {\sqrt 3} u_\uparrow \left[ u_\uparrow s_\downarrow - u_\downarrow s_\uparrow\right] ,& 
\psi_{\Sigma^{*+}} &=& \frac 1 {\sqrt 2} u_\uparrow^2 s_\uparrow , \\
\psi_{\Lambda^0} &=& \frac 1 {\sqrt 2} s_\uparrow \left[ u_\uparrow d_\downarrow - u_\downarrow d_\uparrow\right] ,&
\psi_{\Xi^{*0}} &=& \frac 1 {\sqrt 2} u_\uparrow s_\uparrow^2, \\
\psi_{\Xi^0} &=& \frac 1 {\sqrt 3} s_\uparrow \left[ u_\uparrow s_\downarrow - u_\downarrow s_\uparrow\right] , &
\psi_{\Omega^{-}} &=& \frac 1 {\sqrt 6}  s_\uparrow^3.
\end{array}
\label{12}
\end{eqnarray}
The matrix element of Eq.~(\ref{4}) is computed as
\begin{eqnarray}
\opq^\dagger \sigma^3 \tau^3 \opq \ket{p} &=& \frac 1 {\sqrt 3} u_\uparrow \left[ 3 u_\uparrow d_\downarrow + u_\downarrow d_\uparrow\right],
\label{13}
\end{eqnarray}
on applying Eq.~(\ref{10}) to Eq.~(\ref{12}), so that
\begin{eqnarray}
\me{p}{\opq^\dagger \sigma^3 \tau^3 \opq}{p} &=& \frac 1 3 \left[ 3
\ip{u_\uparrow^2 d_\downarrow}{u_\uparrow^2 d_\downarrow} - 
\ip{u_\uparrow u_\downarrow d_\uparrow}{u_\uparrow u_\downarrow d_\uparrow}  \right] 
\nn \nn
& = &\frac 1 3 \left[ 3\cdot 2 - 1 \right] = \frac 53.
\label{14}
\end{eqnarray}

The $\Theta^+$ pentaquark is an exoticness $E=1$ baryon, and is a $qqq\bar q$
$\mathbf{\overline{10}}_{1/2}$ state in the quark model. The wavefunctions of the $I=I_3$, $J=1/2$ pentaquarks are
\begin{eqnarray}
\psi_{\Theta^+} &=& {1\over 2 \sqrt 3} \left[u_\uparrow d_\downarrow-u_\downarrow d_\uparrow \right]^2 \bar s_\uparrow ,\nn
\psi_{p_{\overline{10}}} &=& {1\over 6} \left[u_\uparrow d_\downarrow-u_\downarrow d_\uparrow \right] \nn
&& \hspace{-1cm} \times \left[ 2 u_\uparrow s_\downarrow \bar s_\uparrow +
 u_\downarrow d_\uparrow \bar d_\uparrow - u_\uparrow d_\downarrow \bar d_\uparrow - 2 u_\downarrow s_\uparrow \bar s_\uparrow \right], \nn
\psi_{\Sigma^+_{\overline{10}}} &=& -{1\over 6} \left[u_\uparrow s_\downarrow-u_\downarrow s_\uparrow \right] \nn
&& \hspace{-1cm} \times \left[ 2 u_\uparrow d_\downarrow \bar d_\uparrow +
 u_\downarrow s_\uparrow \bar s_\uparrow - u_\uparrow s_\downarrow \bar s_\uparrow - 2 u_\downarrow d_\uparrow \bar d_\uparrow \right] ,\nn
\psi_{\Xi_{3/2}^{++}} &=&  -{1\over 2 \sqrt 3} \left[u_\uparrow s_\downarrow-u_\downarrow s_\uparrow \right]^2 \bar d_\uparrow .
\label{10b}
\end{eqnarray}
The wavefunctions of the $I=1$ state $\Theta_1$ in the spin-1/2 $\mathbf{27}$, spin-3/2 $\mathbf{27}$, and the $I=2$ state $\Theta_2$ in the spin-3/2 $\mathbf{35}$ and spin-5/2 $\mathbf{35}$ are
\begin{eqnarray}
\psi(\mathbf{27}_{1/2},\Theta_1)&=&\frac{1}{2 \sqrt{3}} u_\uparrow \left[u_\uparrow d_\downarrow-u_\downarrow d_\uparrow \right] \left[
-u_\uparrow \bar s_\downarrow + u_\downarrow \bar s_\uparrow \right], \nn
\psi(\mathbf{27}_{3/2},\Theta_1)&=& \frac{1}{2 \sqrt{2}}u_\uparrow^2 \bar s_\uparrow \left[u_\uparrow d_\downarrow-u_\downarrow d_\uparrow \right]  ,\nn
\psi(\mathbf{35}_{3/2},\Theta_2)&=& \frac{1}{\sqrt{30}} u_\uparrow^3\left[-
u_\uparrow \bar s_\downarrow + u_\downarrow \bar s_\uparrow \right] ,\nn
\psi(\mathbf{35}_{5/2},\Theta_2)&=& \frac{1}{2 \sqrt{6}} u_\uparrow^4 
\bar s_\uparrow ,
\label{27}
\end{eqnarray}
from which the other wavefunctions can  be generated by applying raising and lowering operators. Matrix elements involving pentaquark states can be computed using Eqs.~(\ref{10b},\ref{27}).

The $qqqq \bar q$ states in the quark model also include a $\mathbf{8}_{1/2}$. This $qqqq \bar q$ octet is a $qqq$ octet plus a flavor singlet $q \bar q$ pair, and has exoticness $E=0$ even though it is a five-quark state. It is useful to have wavefunctions for baryons states containing $q \bar q$ pairs. Define
\begin{eqnarray}
\sing &=& \bar q q = N_- ,
\label{32}
\end{eqnarray}
which annihilates a $\bar qq$ pair, and its hermitian conjugate $\sing^\dagger$ which creates a $\bar qq$ pair. We will need the commutation relations
\begin{eqnarray}
\left[ \sing,\sing^\dagger \right] &=& 2 N_F + N_q + N_{\bar q}, \nn
\left[ \Lambda^A_-,\sing^\dagger \right] &=& \Lambda^A_q-\Lambda^A_{\bar q} ,\nn
\left[ \Lambda^A_q,\sing \right] &=&-\Lambda^A_- , \nn
\left[ \Lambda^A_{\bar q},\sing \right] &=&\Lambda^A_- ,
\label{33}
\end{eqnarray}
where we use the notation of Ref.~\cite{JM}, $\Lambda^A_q = q^\dagger \Lambda^A q$, 
$\Lambda^A_{-} = \bar q \Lambda^A q$, $\Lambda^A_{\bar q} = \bar q^\dagger \bar \Lambda^A \bar q$, where $\Lambda^A,\bar \Lambda^A$ are spin-flavor matrices in the fundamental and anti-fundamental of $SU(6)$.

All minimal states satisfy the constraint that flavor singlet annihilation is impossible, and so obey
\begin{eqnarray}
\sing \ket{ \psi } = 0.
\label{34}
\end{eqnarray}
We now construct a non-minimal $\mathbf{8}$, denoted by $\mathbf{8}^\prime$, and defined by
\begin{eqnarray}
\ket{ \mathbf{8^\prime} } &=&  \lambda\, \sing^\dagger \ket{ \mathbf{8} } ,
\label{35}
\end{eqnarray}
where $\lambda$ is a normalization constant. The value of $\lambda$ is determined by computing the normalization:
\begin{eqnarray}
\ip{ \mathbf{8^\prime} }{ \mathbf{8^\prime} } &=& \abs{\lambda}^2
\me{ \mathbf{8} } {\sing \sing^\dagger} { \mathbf{8} } \nn
&=&  \abs{\lambda}^2
\me{ \mathbf{8} } {\left[\sing, \sing^\dagger\right]} { \mathbf{8} } =  9 \abs{\lambda}^2 ,\nn
\Rightarrow \lambda &=& \frac 1 3.
\label{36}
\end{eqnarray}
In the holomorphic representation
\begin{eqnarray}
\sing^\dagger \to S = Q_{\imath \alpha} \bar Q^{\imath \alpha}.
\end{eqnarray}

The matrix element of an operator such as the Hamiltonian $H$ can be computed similarly:
\begin{eqnarray}
\me{ \mathbf{8^\prime} }{H}{\mathbf{\overline{10}}} = 
\lambda \me{ \mathbf{8} }{\sing H}{\mathbf{\overline{10}}} = \lambda 
\me{ \mathbf{8} }{\left[\sing, H\right]}{\mathbf{\overline{10}}} ,
\label{38}
\end{eqnarray}
and the commutator can be evaluated with the help of Eq.~(\ref{33}). Matrix elements of $\Delta E=0$ operators involving non-minimal states are equal to matrix elements of $\Delta E=-1$ operators involving the corresponding minimal state with the same flavor quantum numbers. Of particular interest are the matrix elements of $T^8$, $T^8_q$ and $T^8_{\bar q}$,
\begin{eqnarray}
\me{ \mathbf{8^\prime} }{T^8}{\mathbf{\overline{10}}} &=& 0 ,\nn
\me{ \mathbf{8^\prime} }{T^8_q}{\mathbf{\overline{10}}} &=& 
\frac 1 3
\me{ \mathbf{8} }{T^8_-}{\mathbf{\overline{10}}} , \nn
\me{ \mathbf{8^\prime} }{T^8_{\bar q}}{\mathbf{\overline{10}}} &=& 
-\frac 1 3
\me{ \mathbf{8} }{T^8_-}{\mathbf{\overline{10}}},
\label{39}
\end{eqnarray}
which are needed to analyze the masses of exotic baryons including $SU(3)$ breaking~\cite{JM4}.


The results so far have been for $N_c=3$. It is  simple to extend the results to arbitrary $N_c$.  Let
\begin{eqnarray}
\dq &=&\opu_\uparrow \opd_\downarrow  - \opu_\downarrow \opd_\uparrow,
\label{15}
\end{eqnarray}
which is a spin and isopspin singlet combination of $u$ and $d$ quarks, with
\begin{eqnarray}
\left[\dq,\dq^\dagger\right] &=& \nud + 2,
\label{16}
\end{eqnarray}
where $\nud$ counts the number of $u$ and $d$ quarks. In the holomorphic representation
\begin{eqnarray}
\dq^\dagger \to \dqh = u_\uparrow d_\downarrow - u_\downarrow d_\uparrow .
\label{19}
\end{eqnarray}
The state
\begin{eqnarray}
\ket{\avar}={(\opu_\uparrow^\dagger)^\avar \over \sqrt{\avar!}} \ket{0},\qquad \psi_{\avar} = {u_\uparrow^\avar \over \sqrt{\avar!}}.
\label{17}
\end{eqnarray}
is a state with $I=J=I_3=J_3=\avar/2$. The state
\begin{eqnarray}
\ket{ n,\avar} = { \left(\dq^\dagger\right)^n  \over  \sqrt {\sigma_{n,\avar}}}\ket{\avar},\quad
\psi_{n,\avar} &=& {\left[u_\uparrow d_\downarrow - u_\downarrow d_\uparrow\right]^n u_\uparrow^\avar \over  \sqrt {\avar! \sigma_{n,\avar} }} .\qquad
\label{18}
\end{eqnarray}
is a baryon with $I=J=I_3=J_3=\avar/2$ and $N_c=2n+\avar$ quarks, and $\sigma_{n,\avar}$ is a normalization constant. One can similarly construct exotic baryons for arbitrary $N_c$ by multiplying the $N_c=3$ wavefunctions by powers of $\dqh^n$.

The normalization $\sigma_{n,\avar}$ is given by
\begin{eqnarray}
\sigma_{n,\avar} &=&
\me{\avar}{\dq^n \left(\dq^\dagger\right)^n}{\avar}.
\end{eqnarray}
This can be evaluated by expanding $\dq^n,\left(\dq^\dagger\right)^n$ in a power series and summing the various terms. An alternate method is to use the commutation relation Eq.~(\ref{16}).
Commuting one $\dq$ past $\left(\dq^\dagger\right)^n$ to annihilate $\ket{\avar}$ gives
\begin{eqnarray}
\dq  \left(\dq^\dagger\right)^n\ket{\avar} &=& \sum_{r=0}^{n-1}
\left(\dq^\dagger\right)^r \left[\dq,\dq^\dagger\right] \left(\dq^\dagger\right)^{n-r-1}
\ket{\avar} \nn
&=& \sum_{r=0}^{n-1}
\left(\dq^\dagger\right)^r \left( \nud +2\right) \left(\dq^\dagger\right)^{n-r-1}
\ket{\avar} \nn
&=& n(n+\avar+1) \left(\dq^\dagger\right)^{n-1} \ket{\avar},
\label{29}
\end{eqnarray}
so that 
\begin{eqnarray}
\sigma_{n,\avar} = n(n+\avar+1) \sigma_{n-1,\avar}.
\end{eqnarray}
Using $\sigma_{0,\avar}=1$ gives
\begin{eqnarray}
\sigma_n = \frac {n! (n+\avar+1)!}{(\avar+1)!} .
\label{23}
\end{eqnarray}

Eq.~(\ref{29}) can be derived more simply using the holomorphic representation, and follows from the chain rule,
\begin{eqnarray}
\dq  \left(\dq^\dagger\right)^n\ket{\avar} &\to& \left( {\partial \over \partial u_\uparrow}{\partial \over \partial d_\downarrow}-{\partial \over \partial u_\downarrow}{\partial \over \partial d_\uparrow} \right) \dqh^n {u_\uparrow^b \over \sqrt{b!}} \nn
&=& n(n+\avar+1) \dqh^{n-1} {u_\uparrow^b \over \sqrt{b!}}.
\label{25alt}
\end{eqnarray}

The matrix element of $\opA^{33}=\opq^\dagger \sigma^3 \tau^3 \opq$ can be computed similarly, using
\begin{eqnarray}
\opA^{33} \ket{\avar} &=& \avar \ket{\avar},
\end{eqnarray}
and
\begin{eqnarray}
\left[\opA^{33},\dq^\dagger\right] &=& \widetilde \dq^\dagger,\nn
\opu_\uparrow \opd_\downarrow  + \opu_\downarrow \opd_\uparrow &=& \widetilde \dq  ,
\nn
\left[\dq, \widetilde \dq^\dagger \right] &=& \opA^{33} .\nn
\label{21}
\end{eqnarray}
Defining
\begin{eqnarray}
\sigma_{n,\avar} \xi_{n,\avar} &=& \me{\avar}{\dq^n\ \opA^{33} \left(\dq^\dagger\right)^n}{\avar}, \nn
\sigma_{n,\avar} \eta_{n,\avar} &=& \me{\avar}{\dq^n\ \widetilde \dq^\dagger \left(\dq^\dagger\right)^{n-1}}{\avar} ,
\label{24}
\end{eqnarray}
and commuting $\opA^{33}$ through in the first equation, and one $\dq$ through in the second, as in Eq.~(\ref{29}) gives the recursion relations
\begin{eqnarray}
\xi_{n,\avar} &=& 2 n \eta_{n,\avar} + \avar , \label{25} \\
n(n+\avar+1) \eta_{n,\avar} &=& (n-1)(n+\avar+2) \eta_{n-1,\avar} + \avar .\nonumber
\end{eqnarray}
The same relations follow more simply in the holomorphic representation using the differential form of the operator, Eq.~(\ref{10}), and the chain rule.

The solution to the recursion relations is
\begin{eqnarray}
\eta_{n,\avar} &=& {\avar \over \avar+2}, \nn
\xi_{n,\avar} &=& {\avar(2n+\avar+2) \over (\avar+2)}={\avar(N_c +2) \over (\avar+2)},
\label{30}
\end{eqnarray}
using $N_c=2n+\avar$.
The matrix element for $\avar=1$ is
\begin{eqnarray}
g_A(NN\pi)= \xi_{n,1} = {N_c +2 \over 3}
\label{41}
\end{eqnarray}
which is the value of the $NN\pi$ axial coupling as a function of $N_c$ in the quark model~\cite{Karl}, and reduces to $5/3$ for $N_c=3$. The matrix element of $\opA^{33}$ in the $\Delta$ as a function of $N_c$ is given by using $\avar=3$, 
\begin{eqnarray}
g_A(\Delta\Delta \pi) = {3(N_c+2)\over5}.
\label{42}
\end{eqnarray}

One can similarly compute the $ N \to \Delta$ transition axial matrix element of
$\opA^{++}$. The operator relations needed are:
\begin{eqnarray}
&& \opA^{++} = \opu_\uparrow^\dagger \opd_\downarrow,\quad
\left[\opA^{++},\dq^\dagger \right] = \left( \opu^\dagger \right)^2 ,\quad
\opA^{++} \ket{\avar} = 0 ,\nn
&& \left( \opu^\dagger \right)^2 \ket{\avar} = \sqrt{(\avar+1)(\avar+2)} \ket{\avar+2}.
\end{eqnarray}
By a method similar to the derivation of Eq.~(\ref{29},\ref{25alt})
\begin{eqnarray}
\me{n-1,\avar+2}{\opA^{++}}{n,\avar} &=& \frac 1 2 \sqrt{{n(\avar+1)(n+\avar+2) \over (\avar+3)}}\nn
 && \hspace{-3cm}= \frac 1 4 \sqrt{{(\avar+1)(N_c-\avar)(N_c+\avar+4) \over (\avar+3)}},\nn
 && \hspace{-3cm}= \frac {N_c+2} 4 \sqrt{\frac{\avar+1}{\avar+3}}
 \sqrt{1-\frac{(\avar+2)^2}{(N_c+2)^2} },
\label{362}
\end{eqnarray}
which for $\avar=1$ reduces to
\begin{eqnarray}
\me{\Delta}{\opA^{++}}{p} &=& \frac 1 4 \sqrt{{(N_c-1)(N_c+5) \over 2}} .
\label{361}
\end{eqnarray}
The results for $g(NN\pi)$, $g(\Delta\Delta\pi)$ and $g(\Delta N\pi)$ have been obtained earlier~\cite{Karl,DJM2,collins}. 
Note the dependence of these quantities on $N_c+2$. In Ref.~\cite{DJM2}, it was shown that quark matrix elements are given as an expansion in  $J^2/(N_c+2)$, and the
results  Eq.~(\ref{41},\ref{42},\ref{362}) agree with the general form given there.

The connection of the quark model to the soliton model is given by constructing spin-flavor coherent states for baryons containing $N_c$ quarks
\begin{eqnarray}
\ket{\eta; N_c} &=& {(\eta^{\imath \alpha} \opq_{\imath \alpha}^\dagger)^{N_c} \over \sqrt{ N_c!}} \ket{0},
\label{37}
\end{eqnarray}
where the coefficient $\eta$ is normalized to $\eta^{\imath \alpha}\eta^*_{\imath \alpha}=1$. The corresponding holomorphic wavefunction is
\begin{eqnarray}
\psi_{\eta; N_c} &=& {(\eta^{\imath \alpha} Q_{\imath \alpha})^{N_c} \over \sqrt{ N_c!}}.
\end{eqnarray}
The chiral soliton in the standard configuration $U_0$, with $(\mathbf{I+J})U_0=0$ corresponds to $\eta^{\imath \alpha}=\eta^{\imath \alpha}_0$, where
\begin{eqnarray}
\eta_0^{\imath \alpha} Q_{\imath \alpha} &=& \frac{1}{\sqrt 2} \left( u_\downarrow - d_\uparrow \right).
\end{eqnarray}
Rotated soliton configurations are given by applying the corresponding spin and flavor rotations to $\eta^{\imath \alpha}_0$.
The inner product of two coherent states is
\begin{eqnarray}
\ip{\eta^\prime; N_c}{\eta; N_c} &=& \left( \eta^{\imath \alpha} \eta^{\prime *}_{\imath \alpha} \right)^{N_c}.
\end{eqnarray}
In the limit $N_c \to \infty$, this reduces to a $\delta$-function in $\eta$. The overlap of a soliton and a rotated soliton also vanishes unless the rotation is the identity transformation, which is why the quark and soliton models give the same matrix elements as $N_c \to \infty$~\cite{Manohar:1984}. The spin-flavor coherent states Eq.~(\ref{37}) were used in Ref.~\cite{collins} to compute the $N_c \to \infty$ matrix elements in terms of simple integrals.

Excited baryons such as the $[\mathbf{70},1^-]$ are not in completely symmetric spin-flavor state, since one of the quarks is in an orbitally excited state. The bosonic operator formulation automatically constructs only completely symmetric states. One can include excited baryons by introducing a new operator $\opq^\prime$ for excited quarks. States generated by $\opq^\dagger$ and $\opq^{\prime \dagger}$ have no symmetry restriction between the $\opq^\dagger$ and $\opq^{\prime \dagger}$ spin-flavor indices. This  allows one to treat the $[\mathbf{70},1^-]$ baryons for arbitrary $N_c$, using a core of $N_c-1$ ground state quarks, and one excited quark~\cite{seventya,seventyb,seventyc,seventyd}. Similarly, one can treat heavy quark baryons by introducing operators $\mathfrak{c}$ and $\mathfrak{b}$ that act on $c$ and $b$ quarks, and constructing states of $N_c-1$ light quarks and one heavy quark~\cite{heavybaryons}. The advantage of using bosonic operators decreases rapidly with the number of quarks that must be treated as special.

\acknowledgments

This work was supported in part by the Department of Energy under Grant 
DE-FG03-97ER40546.

\end{document}